# Spontaneously generated flux ropes in 3-D magnetic reconnection

Shi-Chen Bai[1,2,3], Ruilong Guo[2*], Yuchen Xiao[2], Quanqi Shi[2*], Zhonghua Yao[4], Zuyin Pu[5], Wei-jie Sun[6], Alexander W. Degeling[2], Anmin Tian[2], I. Jonathan Rae[7], Shutao Yao[2], Qiu-Gang Zong[5], Suiyan Fu[5], Yude Bu[1], Christopher T. Russell[8], James L. Burch[9] and Daniel J. Gershman[10]

[1]School of Mathematics and Statistics, Shandong University, Weihai, Shandong, 264209, China

[2]Shandong Provincial Key Laboratory of Optical Astronomy and Solar-Terrestrial Environment, Institute of Space Sciences, Shandong University, Weihai 264209, People's Republic of China

[3]Weihai Institute for Interdisciplinary Research, Shandong University, Weihai, Shandong, 264209, China

[4]Key Laboratory of Earth and Planetary Physics, Institute of Geology and Geophysics, Chinese Academy of Sciences, Beijing 100029, People's Republic of China

[5]School of Earth and Space Sciences, Peking University, Beijing 100871, People's Republic of China

[6]Space Sciences Laboratory, University of California, Berkeley, CA 94720, USA

[7]Department of Mathematics, Physics and Electrical Engineering, Northumbria University, Newcastle, UK

[8]Department of Earth, Planetary, and Space Sciences and Institute of Geophysics and Planetary Physics, University of California, Los Angeles, CA 90095, USA

[9]Southwest Research Institute, San Antonio, TX 78238, USA

[10]NASA Goddard Space Flight Center, Greenbelt, Maryland 20771, USA

email: grl@sdu.edu.cn; sqq@sdu.edu.cn

**Abstract**

Magnetic reconnection is the key to explosive phenomena in the universe. The flux rope is crucial in three-dimensional magnetic reconnection theory and are commonly considered to be generated by secondary tearing mode instability. Here we show that the parallel electron flow moving toward the reconnection diffusion region can spontaneously form flux ropes. The electron flows form parallel current tubes in the separatrix region where the observational parameters suggest the tearing and Kelvin-Helmholtz instabilities are suppressed. The spontaneously formed flux ropes could indicate the importance of electron dynamics in a three-dimensional reconnection region.

**Key points**

1. Ion-scale flux ropes roughly perpendicular to the reconnection X-line are identified near the separatrix.
2. The tearing mode and electron Kelvin-Helmholtz instability cannot explain the flux rope formation well.
3. The cylindrical current and flux rope can be generated spontaneously during reconnection.

**Plain language summary**

Recent investigations of local dynamics near the diffusion region have revealed the crucial role of the ion-scale flux rope in energy conversion, plasma acceleration and plasma transport during magnetic reconnection. However, it remained unclear how the flux rope is formed, especially for those with axes roughly perpendicular to the X-line. Using MMS data, we identified flux ropes (FRs) near the separatrix layer close to the diffusion region. The FRs contain distinct axial directions that are roughly perpendicular to the reconnection X-line. The MMS observation indicates that the secondary tearing-mode instability and electron Kelvin-Helmholtz instability are suppressed. There are no signs of the secondary reconnection either. The parallel currents carried by super-Alfvénic electrons funnelled into the diffusion region near the separatrix are in cylindrical shape perpendicular to the magnetic field. Based on the MMS observations, we propose a new picture that the cylindrical current near the separatrix layer, can generate flux ropes spontaneously.

## 1. Introduction

In a two-dimensional scenario, magnetic reconnection occurs in a thin current sheet, yielding an X-shaped magnetic geometry (*Parker*, 1957; *Sweet*, 1958). Close to the current sheet, ions and electrons demagnetize within several tens of ion initial length ($d_i$), forming ion and electron diffusion regions (EDR) (*Cassak and Shay*, 2007; *Le et al.*, 2013; *Vasyliunas*, 1975). The separatrix layer connects the diffusion region, extends over a long distance and divides the regions of plasma inflow and outflow. This 2-D X-shaped geometry is generally valid locally around the diffusion region within several tens of ion initial length in a three-dimensional (3-D) space environment.

In the past decades, research on the diffusion region has revealed several new aspects of 3-D reconnection processes. The ion-scale flux rope (FR) attracts the most attention (*Bakrania et al.*, 2022; *Drake et al.*, 2006b; *Huang et al.*, 2016). The FRs are helical magnetic structures widely observed in the solar atmosphere and in the vicinities of almost all the planets in the solar system (*Jackman et al.*, 2011; *Russell and Elphic*, 1978; *Vignes et al.*, 2004; *Walker and Russell*, 1985). The formation, evolution, and interaction of ion-scale flux ropes are critical in energy conversion, plasma acceleration and plasma transport in 3-D magnetic reconnection (*Daughton et al.*, 2011; *Drake et al.*, 2006a; *Guo et al.*, 2022; *Huang et al.*, 2019; *Huang et al.*, 2012; *Jiang et al.*, 2021; *Sun et al.*, 2022; *R S Wang et al.*, 2016; *S M Wang et al.*, 2020).

The FRs are traditionally believed to be generated between multiple reconnection lines (*Lee and Fu*, 1985; *J Zhong et al.*, 2013), or generated through secondary tearing-mode instability (*Daughton et al.*, 2006; *Drake et al.*, 2006b) and electron Kelvin-Helmholtz (K-H) instability (*Fermo et al.*, 2012; *Z H Zhong et al.*, 2018) during magnetic reconnection. The axes of FRs produced by these mechanisms are approximately parallel to the X-line. However, a large proportion of the FRs contain distinct magnetic field-aligned currents and cannot be well explained by the previous formation theories (*Smith et al.*, 2024; *Z Wang et al.*, 2023; *Xiao et al.*, 2023; *Yang et al.*, 2022; *Yao et al.*, 2020). The highly tilted flux ropes may be related to the progressive spreading of the reconnection line or non-uniform reconnection at the end of the flux rope (*Hughes and Sibeck*, 1987; *Jiang et al.*, 2023; *Kiehas et al.*, 2012). How to generate the flux ropes is still controversial. However, a localized current flowing parallel to a magnetic field has always been seen before the flux rope generation in the formation process related to the reconnection both in simulation and in the laboratory (*Daughton*

*et al.*, 2011; *Fermo et al.*, 2012; *Gekelman et al.*, 2016; *Ji et al.*, 2023; *Tripathi and Gekelman*, 2010). Furthermore, recent PIC simulation has revealed that a localized current flowing parallel to a magnetic field is a sufficient condition for the formation of a flux rope (*Yoon et al.*, 2024). In this letter, we will show how flux ropes are formed spontaneously without tearing mode instability in a magnetic reconnection region.

**2. Data and methodology**

The data we used in this paper come from four instruments onboard the MMS satellites (*Burch et al., 2016*); the fluxgate magnetometer (FGM; *Russell et al., 2016*), the fast plasma investigation (FPI; *Pollock et al., 2016*), and the electric field double probes (EDP; *Lindqvist et al., 2016*; *Ergun et al., 2016*). Several multi-point analysis methods are applied. The locations of MMS1 are determined using velocity obtained from the Spatio-Temporal Difference method (*Shi et al.*, 2006) integrated over time. Magnetic curvature analysis (*Shen et al.*, 2003) is used to calculate the radius of curvature which should increase inside the flux rope due to the existence of the helical magnetic structure.

The location of the flux rope center is determined following the method outlined by *Yang et al.*, 2022, assuming the 1-D cylindrical symmetry of the flux rope. Under this assumption, magnetic field strength contours form concentric circles, and each spacecraft crosses these contours twice, yielding eight intersection points. The center of the flux rope was determined by minimizing the variance in the distances from $N$ sets of $8N$ points to the flux rope center (see equation 5 in *Yang et al.*, 2022). Each set of eight intersection points corresponding to contour of specific magnetic field strength. We can pinpoint the times and locations of each set of eight points where the observed magnetic field matches the contour of a given field strength. The spatial scale of the flux rope is determined by the distance from its center to the boundary of regions where the axial magnetic field is enhanced.

Additionally, we estimate the azimuthal magnetic field strength $B_{azi}$ induced by current along the flux rope's invariant axis (L direction in our case) and compare to the MMS observation under the assumption of a circular cross-section. Using Ampère's law ($\oint B_{azi} \, dl = \mu_0 \int J_L \, dS$). the current-induced magnetic field $B_{azi}$ can be determined by the radial distance $r$ and the current density $J$ along the axial direction: $B_{azi} = \mu_0 J r/2$. The radial distance $r$ is the distance from the flux rope center to the bipolar peaks of the $B_M$ and $B_N$. The current density $J$ can be estimated by the average value between the peaks of the bipolar fluctuation of the magnetic field.

**3. MMS observation**

Figs. 1a-1i shows the overview of a dayside magnetic reconnection event in LMN coordinates observed by the Magnetospheric Multiscale (MMS) four-spacecraft constellations on 2018-Dec-19 close to an electron diffusion region (EDR). The LMN coordinates for the local frame of the current sheet frame are obtained via the Minimum Variance Analysis (MVA) method (*Sonnerup*, 1998) in interval between 15:04:32 and 15:04:35 UT. The L-direction is approximately the orientation of the reconnecting magnetic component, as well as the outflow direction; the M-direction is the out-of-plane direction, i.e., approximately the reconnection X-line direction; and the N-direction is normal to the current sheet. The reconnecting current sheet crossing occurred from 15:04:33.1 UT to 15:04:33.8 UT (marked by the black rectangle). An EDR is encountered during this interval, which is confirmed based on the following measurements: the reversal of $B_L$ at 15:04:33.3 UT (labelled as T3); electron heating (see Fig. 1b); no ion jet while there is an electron jet (outflow $|V_L| > 600\ km/s$ and EDR out-of-plane flow $|V_M| > 600\ km/s$ in a local current frame) (see Figs. 1c-1d); energy dissipation (see Fig. 1e) and crescent electron distribution (see Fig. 1i) mainly in the M direction. The crossing of the separatrix layers at 15:04:28.8 UT (labelled as T1) was confirmed by significant electron parallel heating (Fig. 1b), a clear boundary in electron pitch angle distribution (Fig. 1f) between the inflow (bidirectional electrons) and outflow (isotropic electrons) regions and enhanced wave activities both below and above electron gyrofrequency ($f_{ce}$, white curve in Fig. 1g) (*Huang et al.*, 2016; *Jiang et al.*, 2022; *Retinò et al.*, 2006). The MMS satellites cross the separatrix again at 15:04:30 UT at a location closer to the EDR, where bi-directional electrons in the inflow region became more isotropic approaching the diffusion region. A flux rope is encountered close to the separatrix layer (labelled as T2, highlighted by the green rectangle in Figure 1) at a distance of ~ 1 $d_i$ away from the EDR. The orientation of FR ([0.2036, -0.4537, 0.8676] in the GSM coordinates) is roughly perpendicular to the reconnection line (along [0.6565, -0.5941, -0.4648] in the GSM coordinates), making it remarkably different from the traditional flux rope generated by tearing-mode

instability (*Drake et al.*, 2006b; *Huang et al.*, 2016).

The trajectory of the MMS constellation which is the relative position from 15:04:27 UT normalized by ion inertial length (1 $d_i$ = 55.84 km for this event)) is shown in Fig. 1h which is obtained by integrating the velocity obtained by the STD method over the same period (*Shi et al.*, 2006). Besides the generally concerned movement in the L and N direction, the satellites have a significant motion in the M-direction. The trajectory relative to the reconnection region is illustrated in Fig.2. The configuration of the observed 3D MR is also demonstrated in Fig. 2. The separatrix layer in 3-D varies in the M-direction and consists of sheet-like structure and flux ropes

The flux rope observed close to the separatrix layer features a distinctive axial direction and intense parallel currents. The FR exhibits enhancement at the $B_L$ component (see Fig. 3a) and a bipolar signature in the $B_M$ and $B_N$ components (green and blue curves in Fig. 3b). An intense current along the background magnetic field (roughly L-direction) peaks at 456.2 nA/m² (Figs. 3c-3d). The electron velocity in the L-direction reaches 313 km/s, more than twice the local Alfvén speed ($V_A$=134 km/s in this event), while ion flow remains weak (~ 17 km/s on average) and nearly constant (Fig. 3e). The bulk velocity of electrons within 19-216 eV (dashed black curve in Fig. 3e) shows high consistency with $V_{eL}$. Therefore, the intense $J_L$ is primarily carried by antiparallel electrons with energies ranging from 19 eV to 216 eV. This is further illustrated in the electron energy-pitch angle spectrum (see Fig. 3g and Fig. 5c). Inside the flux rope, the significant reduction in electron differential energy flux at 0-90 pitch angles (see Fig. 3g) results in high anti-parallel velocity, forming the observed parallel current.

The FR is formed spontaneously by the field-aligned electron flow in the separatrix layer. The spatial scale ($R$) of the flux rope is 1.1 $d_i$, based on the distance to the flux rope center (Fig. 3f). Since the flux rope satisfies the force-free condition ($J_\parallel \gg J_\perp$, see Fig. 2d), we use a flux rope model ($B(r) = B_0 exp(-r^2/a^2)$ *Elphic and Russell*, 1983) to fit the MMS1 observation. Using the field strength and the distance to the flux rope center only from MMS1, the field strength at the flux rope center $B_0$ and the asymptotic helical pitch $\alpha$ are determined ($B_0 = 34\ nT\ a = 129$ km ~ 2R) when the differences in magnetic field strength between the model and observation at different radial distances are minimized. The field strengths observed by the other MMS satellites (colored dotted lines) are in good agreement with the model (colored circles, see Fig. 3h). Under the cylindrical configuration, according to Ampère's law, the azimuthal component of the magnetic field $B_{azi}$ generated by the parallel current can be calculated, which is comparable to the MMS observation ($|B_{azi}|/\sqrt{B_t^2 - B_L^2} = 0.96$, for this case). The $B_t$ and $B_L$ are the total field strength and field strength of the L components at the bipolar peaks of the $B_M$ and $B_N$. The good agreement between MMS observations, flux rope models, and azimuthal magnetic field component calculations implies the validity of the cylindrical axial current and that this intense axial cylindrical current directly produces the flux rope, since both the flux rope model and the calculation of current-induced magnetic field assumes the cylindrical shape of the field-aligned current.

The observed flux rope is not formed by the tearing mode instability and electron Kelvin-Helmholtz (K-H) instability. There is no antiparallel magnetic field component for exciting tearing-mode instability around the observed flux rope (Figs. 4a-4c), nor any electron vortex generated by electron K-H instability (Fig. 4j). Besides, during the first separatrix encountering (blue shaded region in Fig. 4) which is ~3 di from the flux rope along the -M direction (see Fig. 1g), $B_M$ reversal accompanied by intense energy dissipation and an electron jet is detected (see Fig. 1d-1e around T1). However, the MMS observation indicates that the tearing mode instability is stable locally and secondary reconnection may not occur. Based on theory, simulation and observation (*Daughton et al.*, 2011; *Jiang et al.*, 2024; *Liu et al.*, 2013; *Liu et al.*, 2018; *Z H Zhong et al.*, 2018), the unstable range of the tearing mode instability in the separatrix layer (the gray shaded region in Fig. 4h) can be approximately determined with the magnetic field (black and red arrows in Figure 4h) on the two sides of the separatrix layer. All of the currents during the separatrix encounter observed by MMS1-3 (Fig. 4d-4f) fall outside the unstable range. Therefore, secondary reconnection cannot occur. The observed energy dissipation could arise from the parallel electron acceleration in the separatrix (*Egedal et al.*, 2012; *Egedal et al.*, 2015). The magnetic field variation in the blue-shaded region is similar to the flux rope crossing. However, the curvature radius in this region decreases, while the curvature radius increases when crossing the flux rope as shown in the yellow-shaded region. The pile-up region of the magnetic field in the separatrix layer (blue shaded region in Fig. 4) has been reported and considered related to the thinning of the separatrix due to the distant active reconnection region (*Holmes et al.*, 2021).

Additionally, compared to the background inflow region (15:04:32.5 to 15:04:33 UT), the electrons inside the flux rope exhibit no signs of heating or acceleration due to secondary reconnection. The electron temperature remains nearly constant (Figs. 1b and 4g), and there are no accelerated electrons within the flux rope (see comparison between Fig.5c and Fig. 5d). Therefore, the secondary reconnection doesn't appear to be the cause of the observed flux rope either. The electrons with a pitch angle range from 0 to 90 degrees cannot arrive at the flux rope from the -L side to the +L side of the X-line along the field lines which suggests that the field line topology inside the flux rope has changed and connected to the EDR (see comparison between Fig. 5c and Fig. 5d).

## 4. Summary and discussion

In the separatrix layer, the $J_L$ distribution along the M direction in the separatrix layer is non-uniformed which is indicated by the different $J_L$ peaks recorded by different MMS satellites (blue shaded region in Fig. 4d). The non-uniform parallel current distribution will generate pinching force ($J \times B$) along the M or N direction. Once the thermal pressure cannot balance the pinching force, the non-uniform parallel current distribution could be unstable and gradually concentrate the electrons into a cylindrical area to form a cylindrical current. Inside the cylindrical area, the helical magnetic field will eventually be produced and evolve into the flux ropes to reach a stable state. This non-equilibrium flux rope formation process starts from a radially localized parallel current has been recently realized in PIC simulation (*Yoon et al.*, 2024). The thinning and expansion of the separatrix will change the strength of localized parallel current based on MMS observation (*Holmes et al.*, 2021), leading to the non-uniformed $J_L$ distribution near the separatrix. However, the formation of a radially localized parallel current is formed near the separatrix is beyond the scope of this paper, which requires three-dimensional numerical studies in the future.

Additional cases of the flux ropes with axes approximately perpendicular to the X-line near the separatrix layer can be found in Figs. 6-7, which are located close to the EDRs reported by *Webster et al.,* (2018). The parallel currents inside the flux ropes are also carried by the electrons streaming towards the X-line. There is no magnetic field and electron velocity shear near the magnetic flux rope to locally excite tearing modes or electron K-H instability (see Figs. 6a-6b and Figs. 7a-7b). There is good agreement between MMS observation, magnetic flux rope model, and calculation of azimuthal magnetic field components ($|B_{azi}|/\sqrt{B_t^2 - B_L^2} = 0.95$ and 0.82 for these cases). Therefore, the MMS observation in these cases also supports the idea that these flux ropes are produced by the cylindrical parallel current.

Combining the above observational evidence, we suggest that flux ropes can be generated directly from intense parallel currents during reconnection without secondary tearing mode instability. This process is facilitated by electrons streaming towards the X-line, which carry the parallel current along the separatrix. The EDR and background environment can be linked through multiple flux ropes. The flux ropes have been found to be essential in 3-D magnetic reconnection (*Daughton et al.*, 2011; *Lapenta et al.*, 2015). They are closely associated with frequently observed 3-D spiral null points (*Eriksson et al.*, 2015; *Guo et al.*, 2022; *Guo et al.*, 2019; *Olshevsky et al.*, 2015), which are central to addressing key unresolved questions in 3-D magnetic reconnection. The previous theories require numerous secondary tearing mode instabilities to occur in the reconnection to form numerous flux ropes. However, the observation results in this letter imply that the spontaneously formed flux ropes near the separatrix layer can be generated without secondary tearing mode instabilities in the 3-D MR process and the 3-D reconnection region would connect the global environments by bunches of field-aligned current flows.

**Data Availability Statement**

**Acknowledgments**

We thank the entire MMS team and MMS Science Data Center for providing the high-quality data for this study. This work was supported by National Natural Science Foundation of China (Grants No. 42225405, 42274220 and 42275135) and the

Natural Science Foundation of Shandong Province (Grant No. ZR2022QD135 and No. ZR2023JQ016).

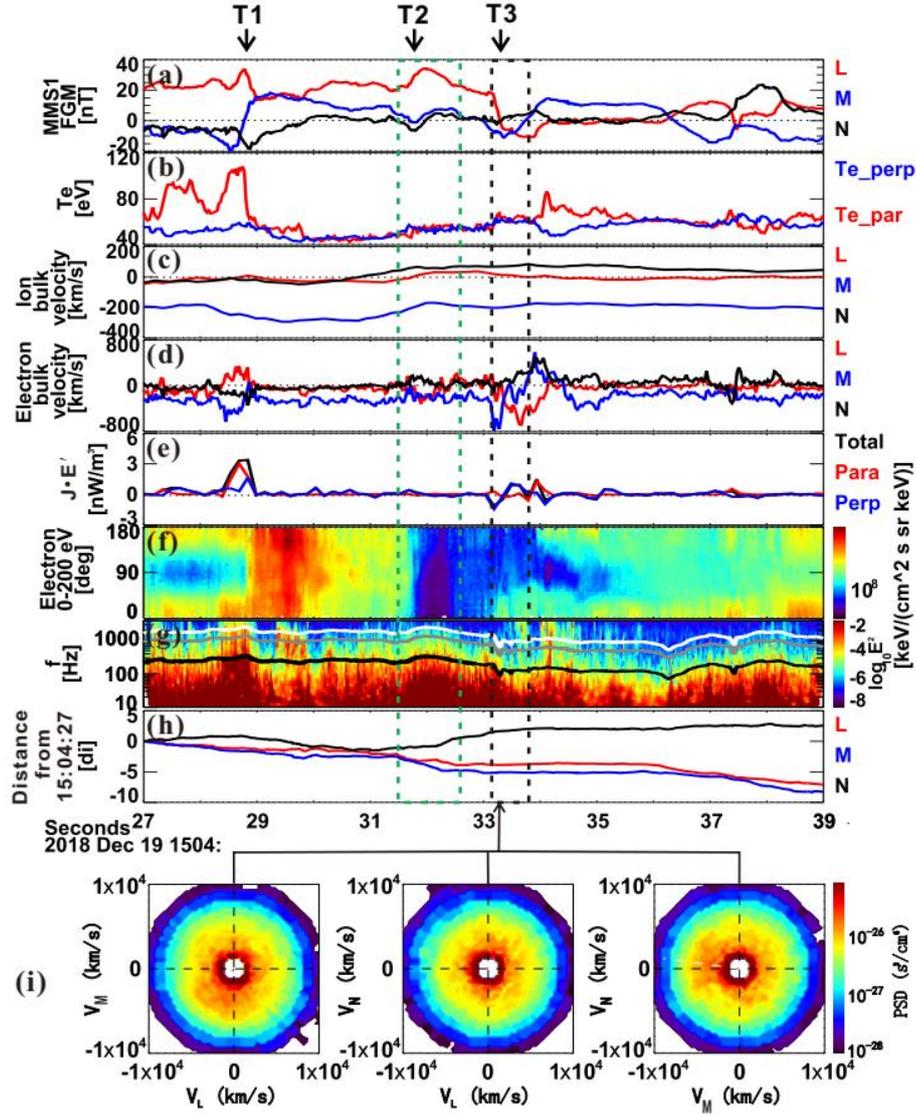

Fig. 1. Overview of MMS1 observations of the dayside reconnection between 15:04:27 to 15:04:39 UT on December 19$^{th}$, 2018. The magnetic field and plasma data are presented in the LMN coordinates by applying minimum variance analysis (MVA) approach. The (x, y, z) GSM components of the L, M, and N axes are L = (0.2036, -0.4537, 0.8676) GSM, M = (0.6565, -0.5941, -0.4648) GSM, and N = (0.7263, 0.6642, 0.1769) GSM. (a) L, M and N components of magnetic field observed by MMS1, (b) parallel and perpendicular electron temperature, (c)-(d) L, M and N components of ion and electron velocity, (e) the energy dissipation of perpendicular and parallel components and the sum of $J \cdot E'$, (f) pitch angle distribution of 0-200 eV electrons, (g) power spectral density of electric field, the white, gray and black curves represent electron gyrofrequency ($f_{ce}$), $0.5 f_{ce}$, $0.1 f_{ce}$, (h) distance of the MMS trajectory in LMN coordinate from 15:04:27 UT, and (i) electron phase space density in L-M, L-N and M-N planes at 15:04:33.3 UT. The flux rope and electron diffusion region are highlighted by green and black rectangles in (a)-(g). The separatrix crossings, the flux rope crossing and the current sheet crossing are labelled as T1, T2 and T3 on the top.

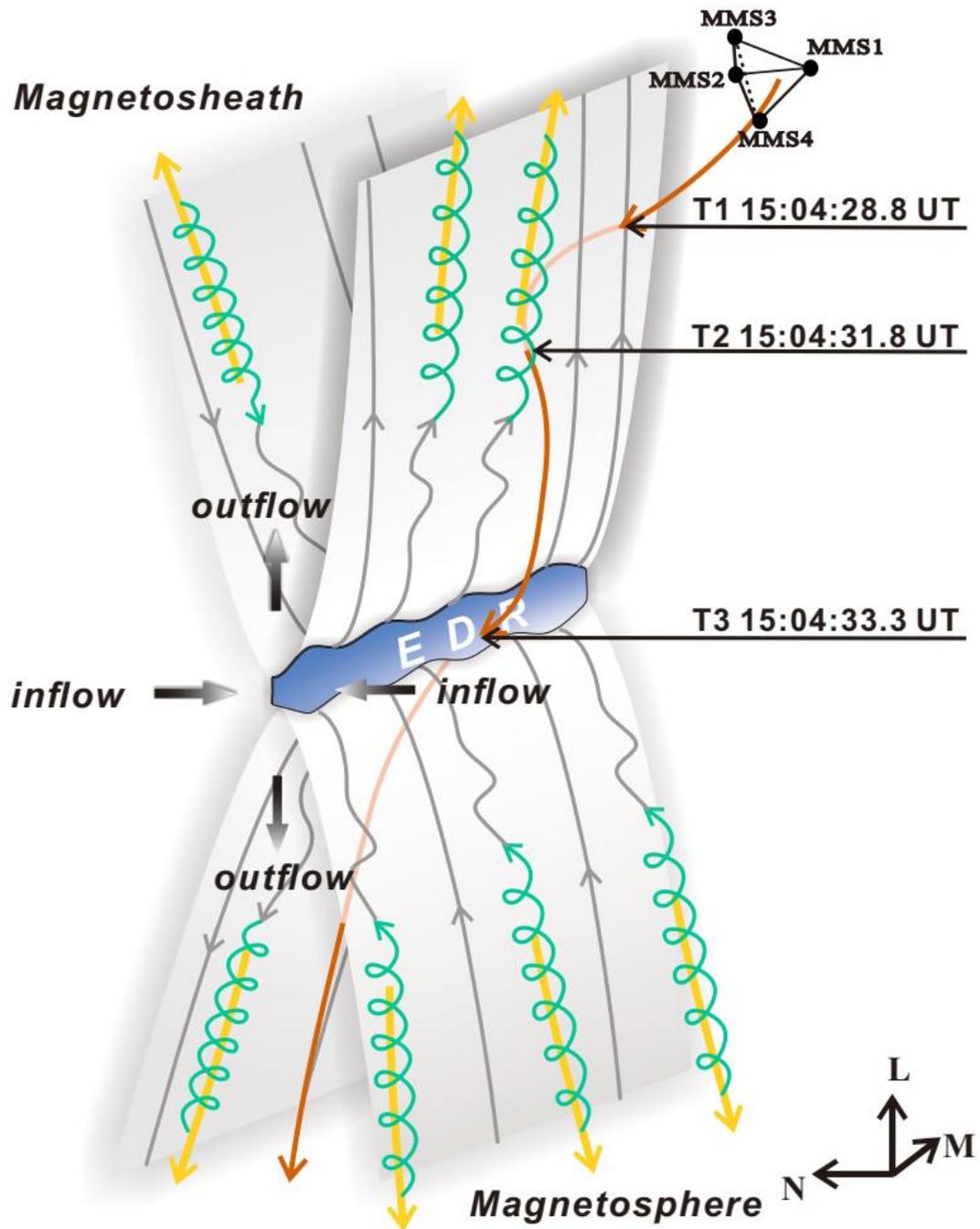

Fig. 2. A schematic of the flux rope perpendicular to the X-line observed in the separatrix layer in three dimensions. The EDR, and reconnection line extends in the M direction. The flux ropes are indicated by green curves. The cylindrical currents are indicated by yellow arrows. The trajectory of MMS is indicated by orange curve.

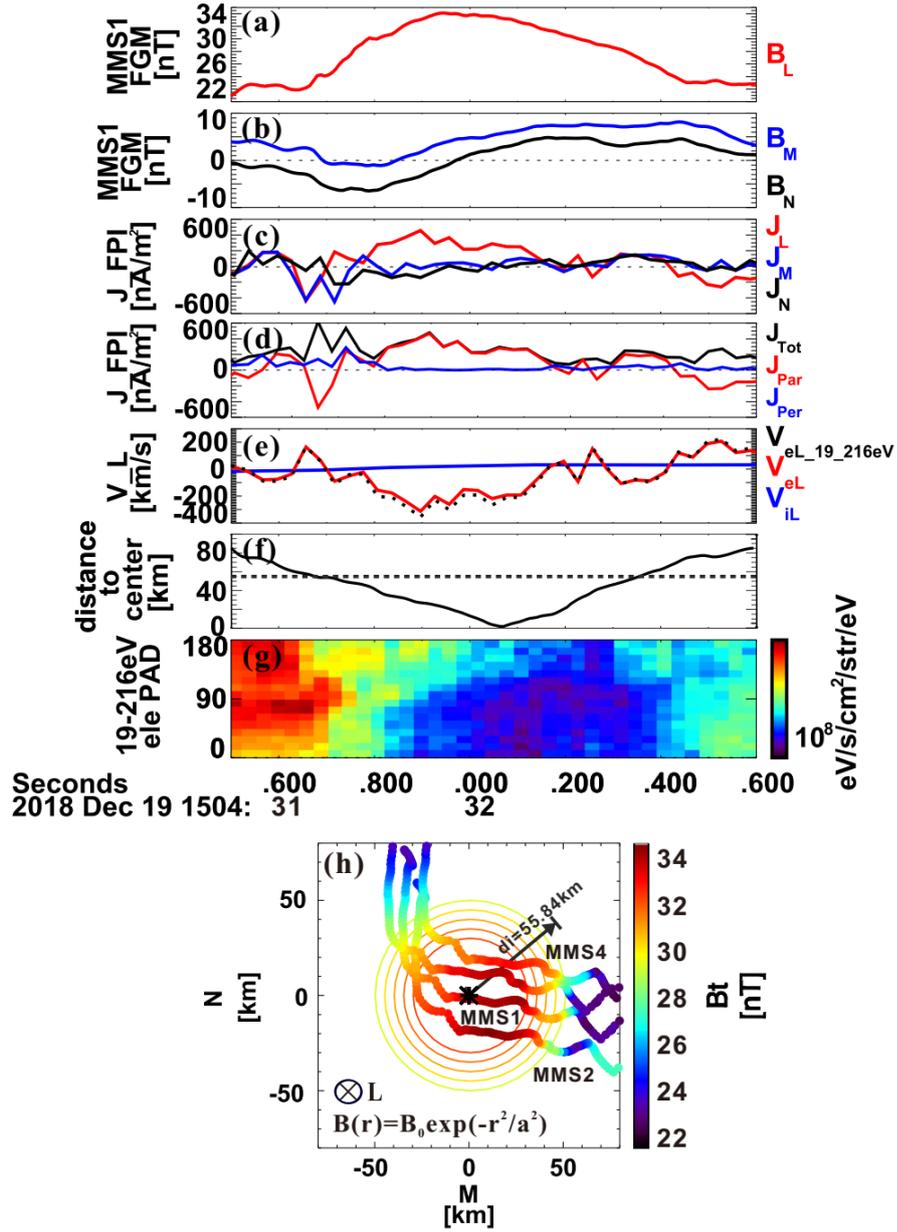

Fig. 3. Detailed observations of the flux rope. (a)-(b) L, M and N components of magnetic field observed by MMS1, (c)-(d) Current density. The red/blue/black curve represents the L/M/N components in (c) and the parallel/perpendicular components and the total current density in (d), (e) L component of ion bulk velocity (black curve), L component of electron bulk velocity (red curve) and L component of the 19-216 eV electron bulk velocity, (f) distance from the MMS1 to the flux rope center, black dashed line indicate the ion inertial length, (g) pitch angle distribution of current-carry electrons, (h) the distribution of the field strength from observation (colored dotted lines) and model (circular colored curves) in the M-N plane. $B_0 = 34\ nT\ a = 129$ for this case.

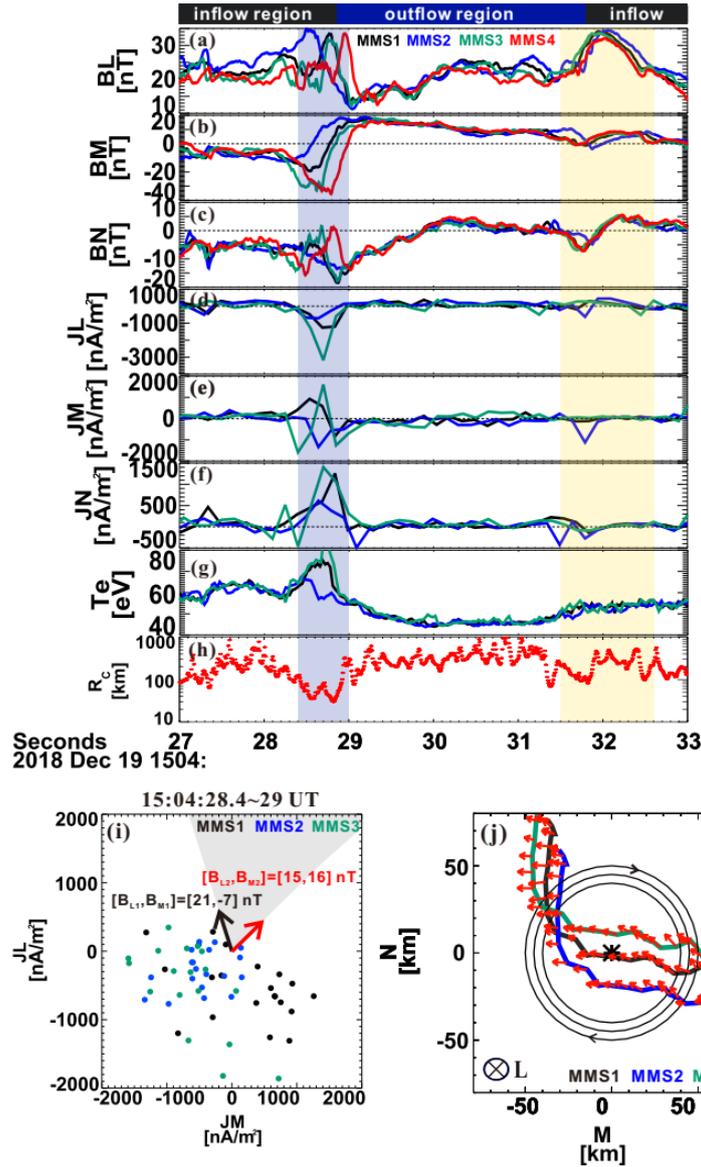

Fig. 4. Four MMS satellite observations of magnetic field and current density in the separatrix layer (a) $B_L$, (b) $B_M$, (c) $B_N$, (d) $J_L$, (e) $J_M$, (f) $J_N$, (g) electron temperature and (h) curvature radius; (i) the current density in the L–M plane $J_{LM}$; the magnetic field in the L–M plane $B_{LM}$ (blue and red arrows) on the two sides of the separatrix defines the unstable range of the tearing instability predicted by the theory (marked by the gray shaded area); (j) the velocity (red arrows) along the MMS 1-3 trajectory in the M-N plane. The blue-shaded region in (a)-(h) highlights the interval of antiparallel $B_M$ components in the separatrix layer. The flux rope is highlighted by the yellow-shaded region in (a)-(h). MMS1/MMS2/MMS3/MMS4 observations are presented by black/blue/green/red curves or dots. The electron data from MMS4 is not available for this case.

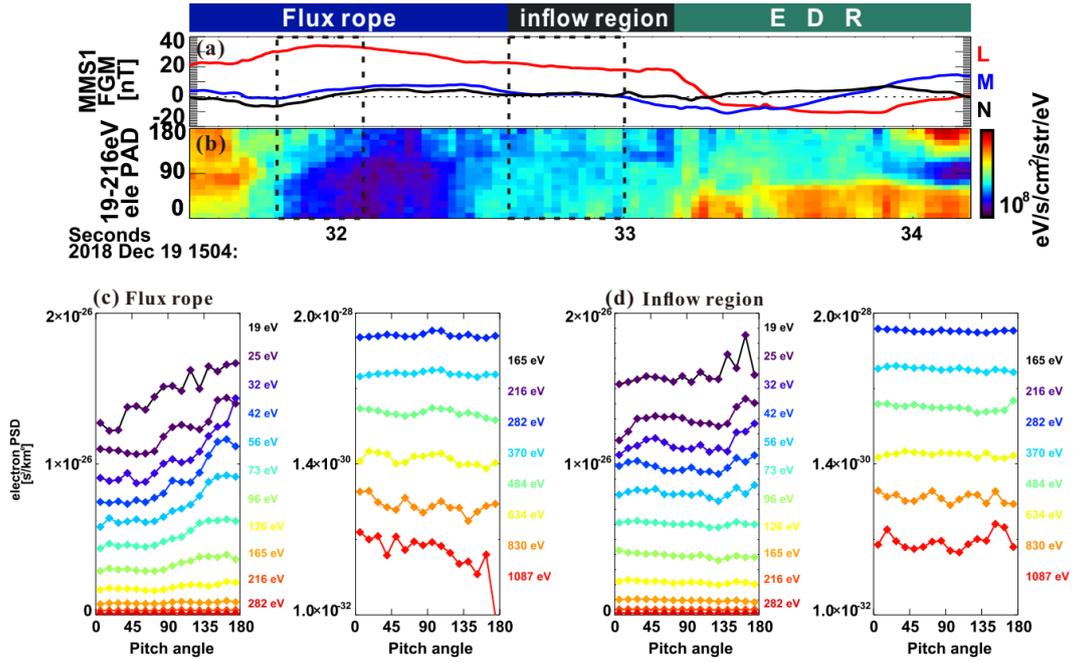

Fig. 5. Comparison of electron pitch angle distribution. (a) L, M and N components of magnetic field observed by MMS1; (b) pitch angle distribution of 19-216 eV electrons; electron pitch angle distribution inside the (c) flux rope and in the (d) inflow region. The time interval of the data used in (c) and (d) are marked by black rectangles in (a) and (b).

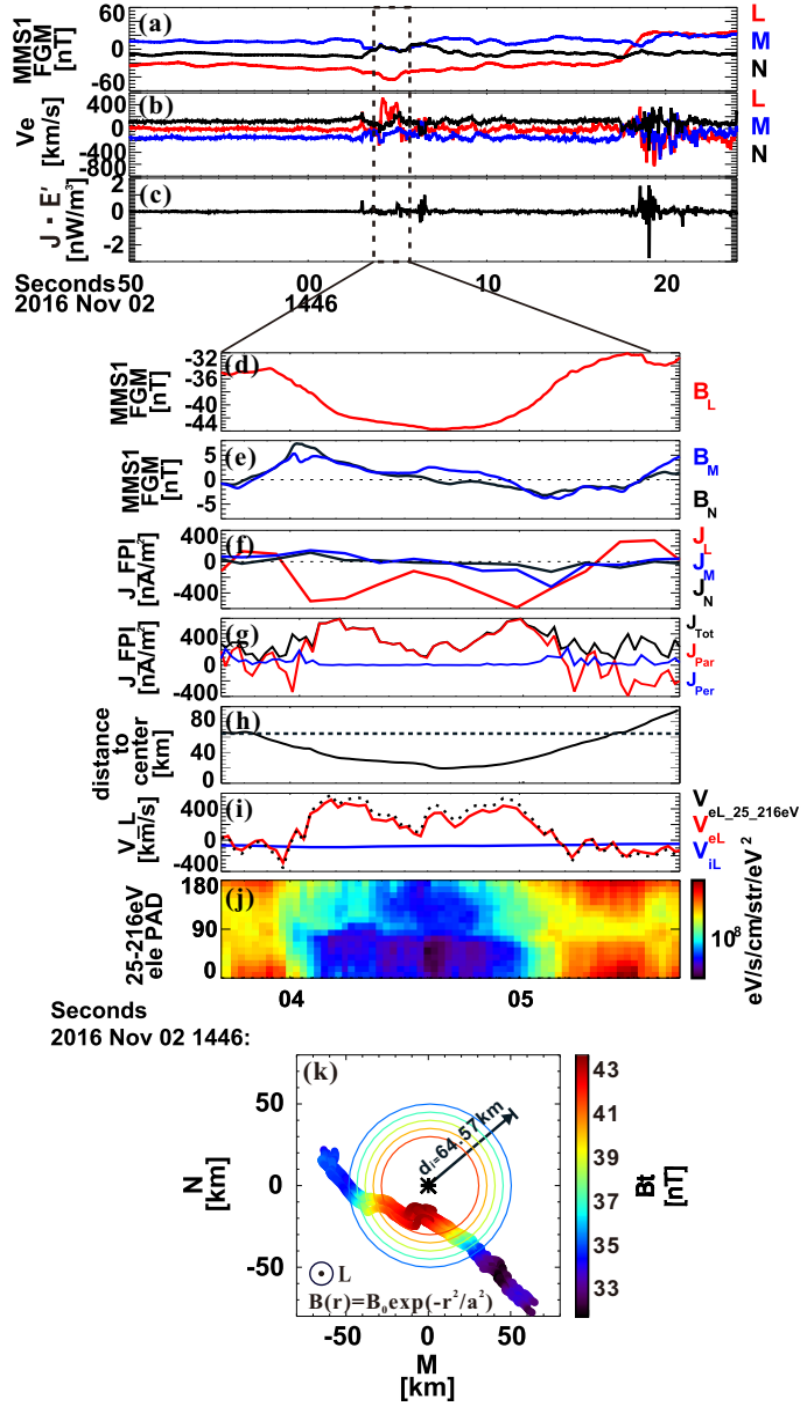

Fig. 6. Overview of MMS observations of the dayside reconnection between 14:45:50 to 14:46:24 UT on November 2$^{rd}$, 2016. (a) L, M and N components of magnetic field observed by MMS1, (b) parallel and perpendicular electron temperature, (b) L, M and N components of electron velocity, (c) the energy dissipation $J \cdot E'$; (d)-(k) Detailed observations of the flux rope. Same format as Fig. 2. $B_0 = 46\ nT\ a = 99\ km$ for this case. The (x, y, z) GSM components of the L, M, and N axes are L = (-0.0750, 0.5824, 0.8094) GSM, M = (0.9577, -0.1840, -0.2212) GSM, and N = (0.2778, 0.7918, -0.5440) GSM. The LMN coordinates is obtained based on magnetic field observations from 14:46:12 to 14:46:24 UT.

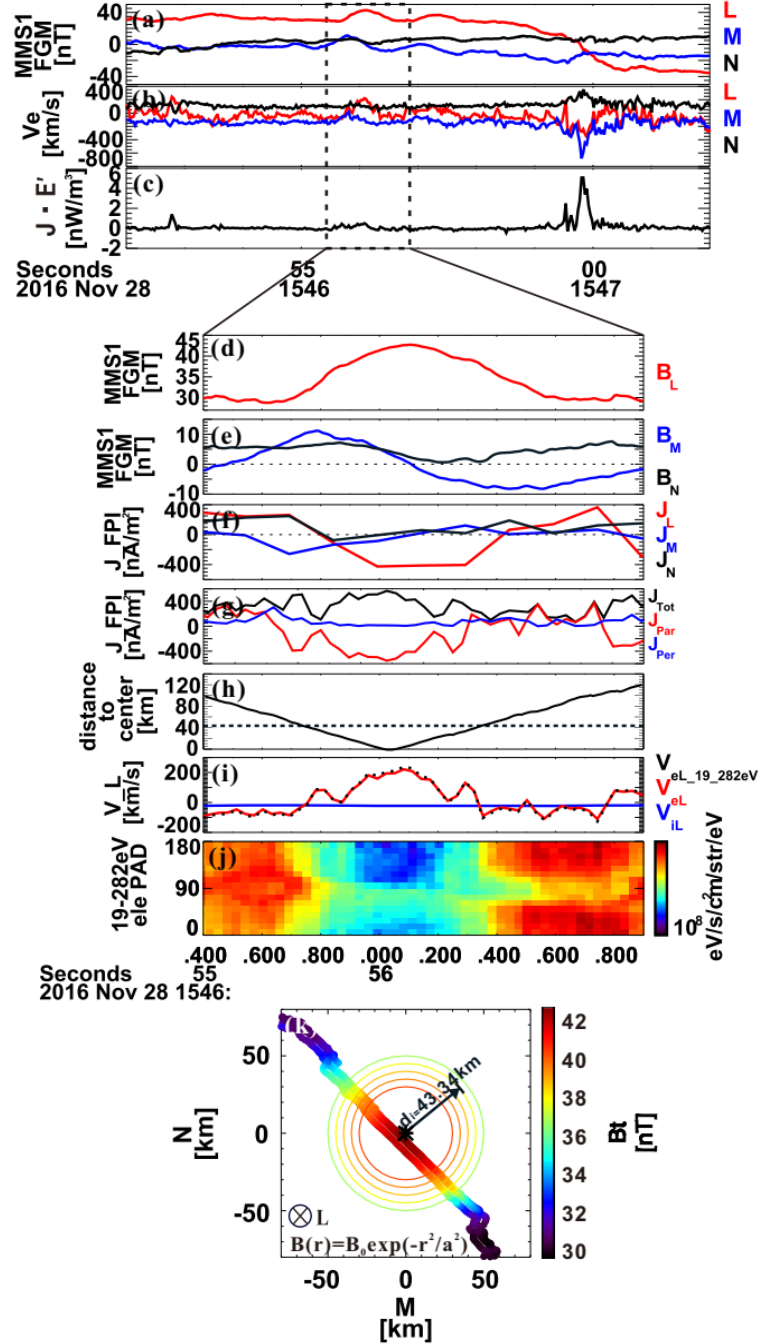

Fig. 7. Overview of MMS observations of the dayside reconnection between 15:46:52 to 15:47:02 UT on November 28$^{th}$, 2016. (a) L, M and N components of magnetic field observed by MMS1, (b) parallel and perpendicular electron temperature, (b) L, M and N components of electron velocity, (c) the energy dissipation $J \cdot E'$; (d)-(k) Detailed observations of the flux rope. Same format as Fig. 2. $B_0 = 43\ nT\ a = 130\ km$ for this case. The (x, y, z) GSM components of the L, M, and N axes are L = (-0.0442, -0.3665, 0.9294) GSM, M = (0.9839, -0.1771, -0.0231) GSM, and N = (0.1730, 0.9134, 0.3685) GSM. The LMN coordinates is obtained based on magnetic field observations from 15:46:58 to 15:47:02 UT.